\input{aipcheck}

\documentclass[
  ,draft            
  ]
  {aipproc}

\layoutstyle{6x9}

\begin{document}

\title{A study of polarization buildup by spin filtering}

\classification{11.80.-m, 13.88.+e, 24.70.+s, 25.43.+t, 29.27Hj}
\keywords      {Antiproton polarization, polarization, antiprotons, spin filtering}

\author{D.~S.~O'Brien}{
  address={School of Mathematics, Trinity College Dublin, Ireland}
}


\begin{abstract}

Many sets of polarization evolution equations have been suggested to describe the method of polarization buildup by spin filtering in storage rings.  In this paper a generic system of polarization evolution equations describing spin filtering is derived and solved, then we compare and contrast this system to other descriptions of spin filtering appearing in the literature.  This is of interest to projects planning to produce a polarized antiproton beam by spin filtering, and to any project utilizing spin filtering in storage rings.  The physical processes responsible for spin filtering are analysed and their contributions to the dynamics of polarization buildup are highlighted.  It is hoped that this will shed light on some of the confusion in the literature.
\end{abstract}

\maketitle


\section{Introduction}
\label{sec:Introduction}

Spin filtering was originally proposed by P.~L.~Csonka in 1968 \cite{Csonka:1968} and was demonstrated experimentally for polarized protons in 1993 by the FILTEX experiment at the Test Storage Ring in Heidelberg \cite{Rathmann:1993xf}.  The spin filtering method is at the heart of the recent PAX (Polarized Antiproton eXperiments) proposal to generate a polarized beam of antiprotons in the HESR ring of FAIR at GSI Darmstadt \cite{Barone:2005pu,Rathmann:2004pm}, the principal aim of which is to investigate the transversity distribution of quarks inside nucleons \cite{Anselmino:2004ki,Anselmino:2007fs}.

Many sets of polarization evolution equations have been suggested to describe the method of polarization buildup by spin filtering.  A generic system is derived in this paper, then this system is compared and contrasted to other descriptions of spin filtering appearing in the literature.  The need for this work was highlighted by H.~O.~Meyer in his summary talk at the Polarized Antiproton Beams - How? Workshop \cite{Meyer:2007} and some of his ideas and suggestions are addressed in this paper.

The spin filtering method of polarization buildup \cite{Csonka:1968,Rathmann:1993xf,Rathmann:2004pm} consists of a circulating beam repeatedly interacting with a polarized internal target in a storage ring.  Many particles are scattered at small angles but remain in the beam.  This introduces a characteristic laboratory acceptance angle $\theta_\mathrm{\,acc}$, scattering above which causes particles to be lost from the beam.  There is also a minimum laboratory scattering angle $\theta_\mathrm{\,min}$, corresponding to the Bohr radius of the atoms in the target, below which scattering is prevented by Coulomb screening.  The two physical processes that contribute to polarization buildup by spin filtering are: (a) spin selective scattering out of the beam, and (b) selective spin-flip.  Thus particles in one spin state may be scattered out of the beam, or have their spin-flipped while remaining in the beam, at a higher rate than particles in the other spin state.  Thus over time one spin state is depleted more than the other leading to a beam polarization.

\section{Polarization evolution equations}
\label{sec:Polarization_evolution_equations}

The two physical processes that contribute to polarization buildup by spin filtering are shown in Figure~1.  The number of particles in the \lq spin up' state can change by three means: (1) \lq spin up' particles being scattered out of the beam, the cross-section for which is labeled as $\sigma^\mathrm{\,out}_+$, (2) \lq spin up' particles being flipped to \lq spin down' particles while remaining in the beam, the cross-section for which is labeled as $\sigma_{+-}$, and (3) \lq spin down' particles being flipped to \lq spin up' particles while remaining in the beam, the cross-section for which is labeled as $\sigma_{-+}$.  Mechanisms (1) and (2) constitute a decrease in the number of \lq spin up' particles and (3) constitutes an increase in the number of \lq spin up' particles.  Correspondingly the number of particles in the \lq spin down' state can also change by three means: (1) \lq spin down' particles being scattered out of the beam, the cross-section for which is labeled as $\sigma^\mathrm{\,out}_-$, (2) \lq spin down' particles being flipped to \lq spin up' particles while remaining in the beam ($\sigma_{-+}$) and (3) \lq spin up' particles being flipped to \lq spin down' particles while remaining in the beam ($\sigma_{+-}$). 

\begin{figure}[h]
\setlength{\unitlength}{1cm}
\begin{minipage}[h]{14.7cm}
{\footnotesize
{\bf FIGURE 1.}  These diagrams describe the two physical processes, selective scattering out of the beam and selective spin-flip, that contribute to polarization buildup by spin filtering in a storage ring.  Black squares represent particles in the \lq spin up' state and white squares represent particles in the \lq spin down' state, while the large grey box represents a polarized target.  In both cases the beam is initially unpolarized with equal numbers of particles in the \lq spin up' and \lq spin down' states.  We label the cross-sections for particles in the \lq spin up' and \lq spin down' states to be scattered out of the beam as $\sigma_+^\mathrm{\,out}$ and $\sigma_-^\mathrm{\,out}$ respectively, the cross-section for a particle in the \lq spin up' state to be flipped to the \lq spin down' state while remaining in the beam as $\sigma_{+-}$ and the cross-section for a particle in the \lq spin down' state to be flipped to the \lq spin up' state while remaining in the beam as $\sigma_{-+}$.  In order for each of these processes to contribute to beam polarization buildup one must have $\sigma_+^\mathrm{\,out} \, \neq \, \sigma_-^\mathrm{\,out}$ and $ \sigma_{+-}\, \neq \, \sigma_{-+}$ respectively.
\\

\begin{minipage}{7.2cm}
\includegraphics[width=6.9cm]{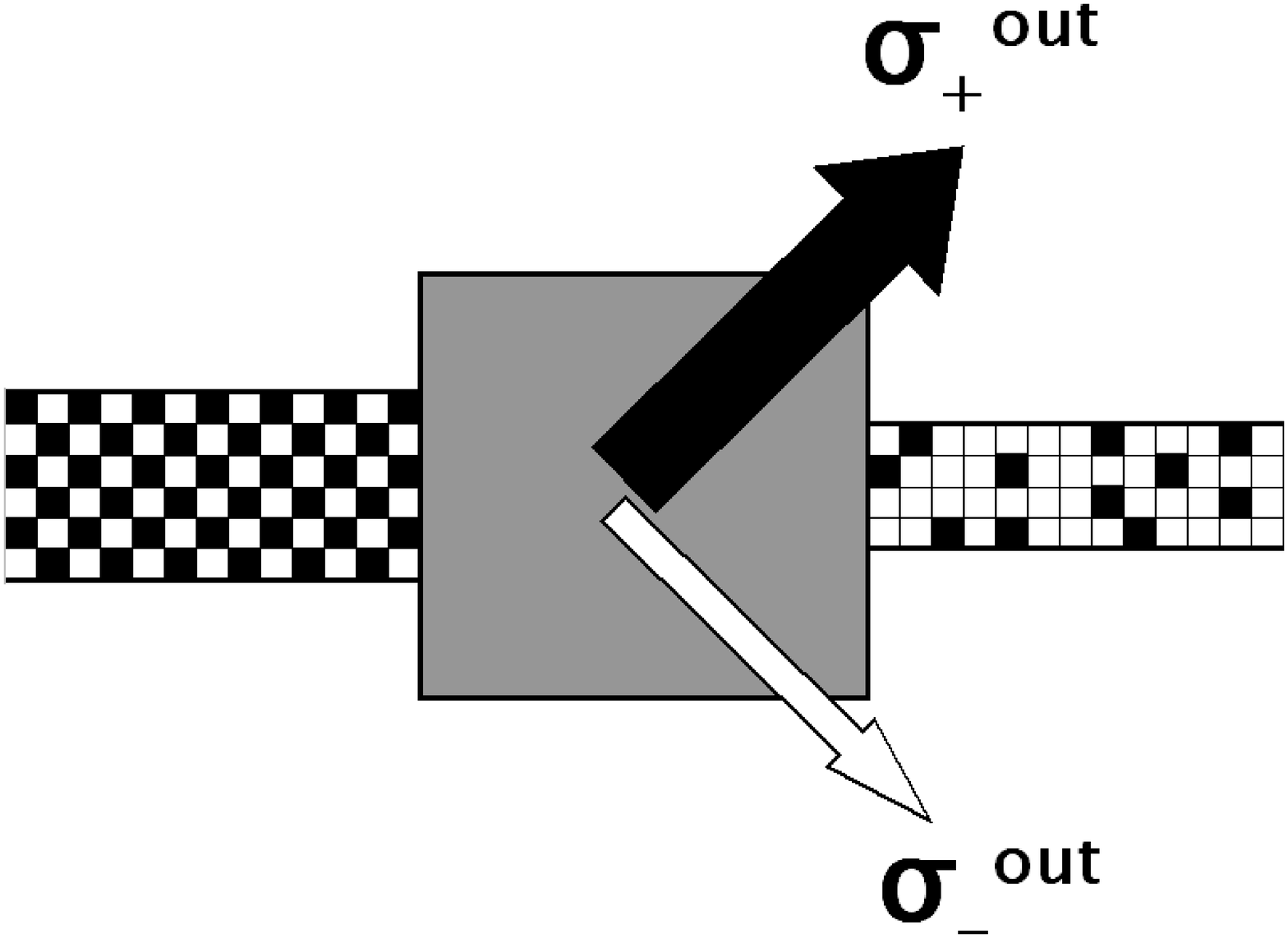}
\end{minipage}
\hfill
\begin{minipage}{7.2cm}
\includegraphics[width=6.9cm]{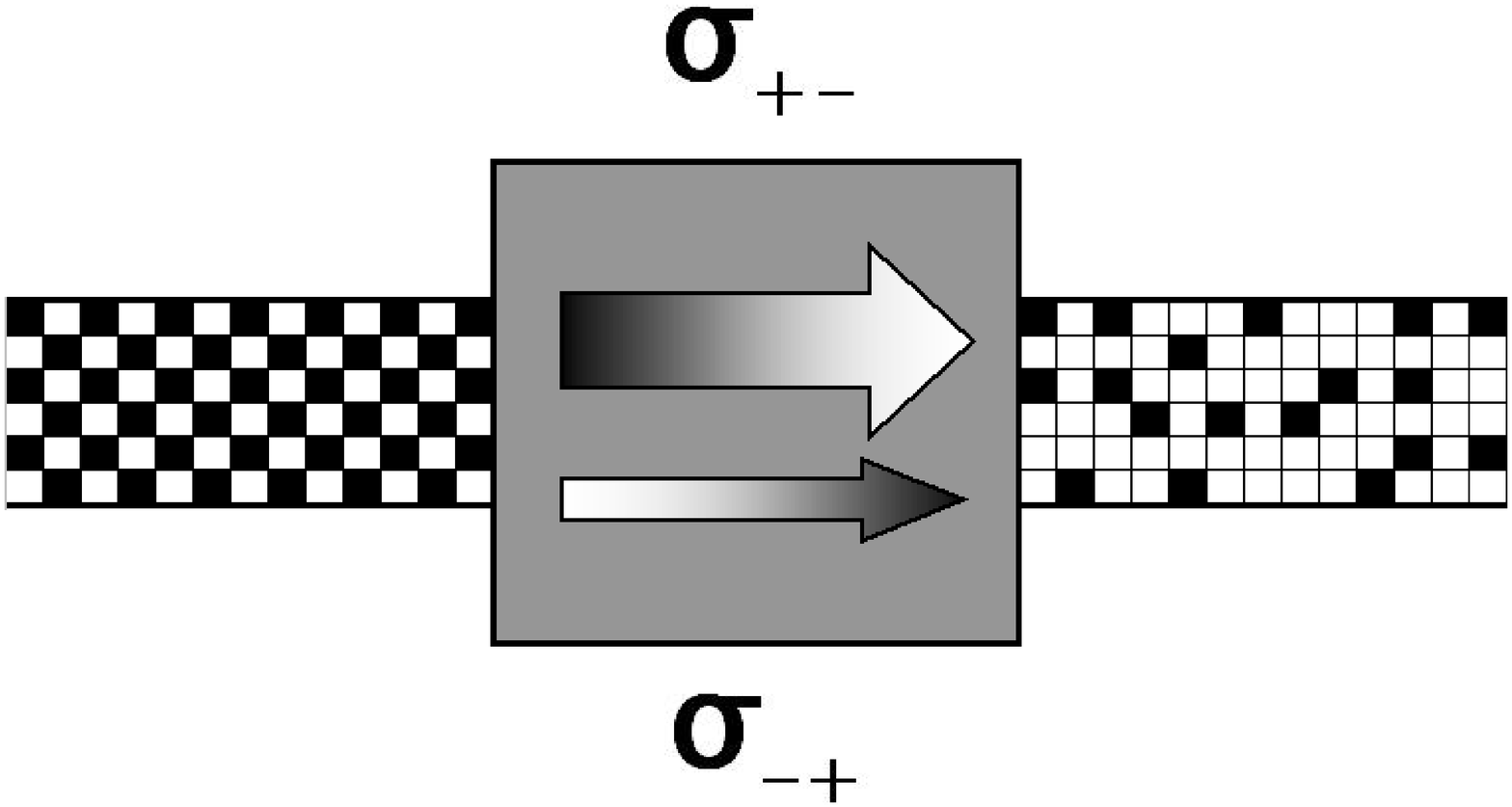}
\end{minipage}
{\bf (A)} The first diagram which represents {\bf selective scattering out of the beam} can be explained as follows:  When interacting with the polarized target at certain energies particles in the \lq spin up' state are scattered out of the beam at a higher rate than particles in the \lq spin down' state, hence the larger black arrow than white arrow.  Thus one is left with a beam that has more particles in the \lq spin down' state, {\it i.e.}\  the beam is now polarized, represented by the excess of white squares in the final beam.  Note that since particles have been scattered out of the ring there are less particles in the beam after interaction than were in the beam initially; this is represented by the smaller final beam.  If the target was unpolarized, particles in both spin states would be lost at equal rates, thus no polarization buildup would occur via this process. \hspace{5em}  {\bf (B)} The second diagram which represents {\bf selective spin-flip} can be explained as follows:  On interaction with the polarized target at certain energies the \lq spin up' to \lq spin down' spin-flip cross-section is larger than the \lq spin down' to \lq spin up' spin-flip cross-section.  This is represented by different size arrows with colors fading from black to white and from white to black respectively.  Thus after interaction with the target the beam will have more particles in the \lq spin down' state than in the \lq spin up' state, {\it i.e.}\  the beam is now polarized, represented by the excess of white squares in the final beam.  Note that the beam intensity is the same after interaction with the polarized target in this process since particles are not lost, they are just flipped from one spin state to the other. If the target was unpolarized, particles in both spin states would have their spins flipped at equal rates, thus no polarization buildup would occur via this process.}
\end{minipage}
\label{fig:Spin_Filtering_Diagrams_With_Symbols}
\end{figure}
All of this can be expressed by the following system of polarization evolution equations \cite{Meyer:2007}\,:
\begin{eqnarray}
\label{eq:N+N-Polarization_evolution_Equations}
\frac{d}{d\,\tau}\left[\begin{array}{c} N_+ \\[2ex] N_-\end{array}\right] & = & - \,n\,\nu \ \left[\begin{array}{ccc} \sigma^\mathrm{\,out}_+ \ + \ \sigma_{+-} & -\,\sigma_{-+}\\[2ex] - \,\sigma_{+-} & \sigma^\mathrm{\,out}_- \ +\  \sigma_{-+}\end{array}\right] \ \left[\begin{array}{c} N_+ \\[2ex] N_-\end{array}\right] \, ,
\end{eqnarray}
where $\tau$ is the time variable, $n$ is the areal density of the target, $\nu$ is the revolution frequency of the beam and $N_+(\tau)$ and $N_-(\tau)$ are the number of beam particles in the \lq spin up' and \lq spin down' states at time $\tau$ respectively.  This set of polarization evolution equations, presented by H.~O.~Meyer \cite{Meyer:2007}, are similar to those treated by A.~I.~Milstein and V.~M.~Strakhovenko \cite{Milstein:2005bx} (note that their $\Omega_{+-}$ is $\sigma_{-+}$ here etc.), W.~W.~MacKay and C.~Montag \cite{MacKay:2006} and V.~F.~Dmitriev, A.~I.~Milstein and V.~M.~Strakhovenko \cite{Dmitriev:2007ms}.

For a beam that is initially unpolarized one imposes the following initial conditions
\begin{equation}
N_+(0) \ = \ N_-(0) \ = \ \frac{N_0}{2}  \, ,
\end{equation}
where $N_0 \,=\, N_+(0) \,+\, N_-(0)$ is the total number of particles in the beam initially.  We define the {\it beam intensity} $N(\tau) \,=\, N_+(\tau) \, + \, N_-(\tau)$ as the total number of particles in the beam at time $\tau$, and the {\it beam total spin} as $J(\tau) \,=\, N_+(\tau) \, - \, N_-(\tau)$ so that polarization of the beam at time $\tau$ is simply given by
\begin{equation}
P(\tau) \ = \ \frac{N_+(\tau) \, - \, N_-(\tau)}{N_+(\tau) \, + \, N_-(\tau)} \ = \ \frac{J(\tau)}{N(\tau)} \, .
\end{equation}
The beam intensity is a physically important variable.  One requires a beam intensity as large as possible to ensure high luminosity in an experiment, which implies greater data taking leading to more accurate results.  A problem with spin filtering where particles are scattered out of the beam is that while the beam polarization increases the beam intensity decreases.

Note some treatments of spin filtering investigate a scenario where no particles are scattered out of the beam, {\it i.e.}\ the maximum scattering angle for the process is less than the ring acceptance angle, which is the case for antiprotons scattering off electrons in an atomic target \cite{Milstein:2005bx,Meyer:1994,Horowitz:1994,Nikolaev:2006gw,O'Brien:2007hu} and for antiprotons scattering off a co-moving beam of electrons or positrons \cite{Walcher:2007sj}.  In these scenarios only selective spin-flip can contribute to polarization buildup, and one avoids the problem of decreasing beam intensity.  The low density of the targets currently available causes the rate of polarization buildup using these methods to be slow, but the enhanced cross-sections at low energies suggested in refs.~\cite{Walcher:2007sj,Arenhoevel:2007} may compensate this difficulty.  We analyze such systems later in the paper.

\pagebreak
\noindent
Before solving this system of polarization evolution equations we shall prove four short results providing a consistency check that the equations accurately describe the physical phenomena we wish to investigate.  This also provides a chance to highlight the dynamical properties of the physical system, which any description of spin filtering must satisfy.
\\
\\
{\bf 1)} {\it If $\sigma^\mathrm{\,out}_+ \, = \, \sigma^\mathrm{\,out}_-$ and $\sigma_{+-}\, = \, \sigma_{-+}$ there will be no buildup of beam polarization, but there will still be loss of beam intensity $N(\tau)$.}
\\
This can be shown as follows: When $\sigma^\mathrm{\,out}_+ \, = \, \sigma^\mathrm{\,out}_-$ and $\sigma_{+-}\, = \, \sigma_{-+}$ the polarization evolution equations presented in eqs.~(\ref{eq:N+N-Polarization_evolution_Equations}) reduce to
\begin{eqnarray}
\frac{d}{d\,\tau}\left[\begin{array}{c} N_+ \\[2ex] N_-\end{array}\right] & = & - \,n\,\nu \ \left[\begin{array}{ccc} \sigma^\mathrm{\,out}_+ \ + \ \sigma_{+-} & -\,\sigma_{+-}\\[2ex] - \,\sigma_{+-} & \sigma^\mathrm{\,out}_+ \ +\  \sigma_{+-}\end{array}\right] \ \left[\begin{array}{c} N_+ \\[2ex] N_-\end{array}\right] \, ,
\end{eqnarray}
{\it i.e.}\
\begin{eqnarray}
\label{eq:N+N-No_Loss_System2}
\frac{d\,N_+}{d\,\tau} & = & - \,n\,\nu \ \left[\,\left(\,\sigma^\mathrm{\,out}_+ \,+\sigma_{+-}\,\right)\,N_+ \ - \ \sigma_{+-}\,N_-\,\right]  \, ,\nonumber \\[2ex]
\frac{d\,N_-}{d\,\tau} & = & - \,n\,\nu \ \left[\,-\,\sigma_{+-}\,N_+ \ + \ \left(\,\sigma^\mathrm{\,out}_+ \,+\, \sigma_{+-}\,\right)\,N_-\,\right] \, ,
\end{eqnarray}
which can be added and subtracted to give
\begin{eqnarray}
\label{eq:System_with_Unpolarized_target1}
\frac{d\,N(\tau)}{d\,\tau} & = &  \, - \, n \, \nu \,\sigma^\mathrm{\,out}_+ \,N(\tau)  \, ,\nonumber \\[2ex]
\frac{d\,J(\tau)}{d\,\tau} & = &  \, - \, n \, \nu \,\left(\,\sigma^\mathrm{\,out}_+ \, + \,2\,\sigma_{+-} \,\right)\,J(\tau) \, ,
\end{eqnarray}
two uncoupled first order separable Ordinary Differential Equations (ODE) which can be integrated to give the solutions
\begin{eqnarray}
\label{eq:System_with_Unpolarized_target2}
N(\tau) & = & N(0)\ \displaystyle{e^{\,-\,n\,\nu\,\sigma^\mathrm{\,out}_+\,\tau}} \ = \ N_0\ \displaystyle{e^{\,-\,n\,\nu\,\sigma^\mathrm{\,out}_+\,\tau}}  \, ,\nonumber\\[2ex]
J(\tau) & = & J(0)\ \displaystyle{e^{\,-\,n\,\nu\,\left(\,\sigma^\mathrm{\,out}_+ \, + \,2\,\sigma_{+-} \,\right)\,\tau}}  \, .
\end{eqnarray}
One sees that $N(\tau)$ will decrease exponentially and $J(\tau)$ which is zero initially will always remain zero, {\it i.e.}\  if $J(0) \,=\, 0$ then $J(\tau) \,=\, 0$ for all $\tau$.  Thus there will be no polarization buildup.  Also in the case when the beam is initially polarized ($J(0) \, \neq \, 0$) its polarization will decrease exponentially to zero, remembering the cross-sections are positive quantities.  We later show that $\sigma^\mathrm{\,out}_+ \, = \, \sigma^\mathrm{\,out}_-$ and $\sigma_{+-}\, = \, \sigma_{-+}$ when the internal target is not polarized.  Thus there will be no polarization buildup by spin filtering if the internal target is unpolarized.

\pagebreak
\noindent
{\bf 2)} {\it If $\sigma^\mathrm{\,out}_+ \, = \, \sigma^\mathrm{\,out}_- \, = \,0$ there will be no loss of beam intensity $N(\tau)\,=\, \mbox{Constant} \,=\, N_0$\,, but there may still be beam polarization buildup.}
\\
This can be shown as follows: When $\sigma^\mathrm{\,out}_+  \,=\, \sigma^\mathrm{\,out}_- \, = \,0$ (this happens when the maximum scattering angle for the process is less than the ring acceptance angle) the polarization evolution equations presented in eqs.~(\ref{eq:N+N-Polarization_evolution_Equations}) reduce to
\begin{eqnarray}
\label{eq:N+N-No_Loss_System}
\frac{d}{d\,\tau}\left[\begin{array}{c} N_+ \\[2ex] N_-\end{array}\right] & = & - \,n\,\nu \ \left[\begin{array}{ccc} \sigma_{+-} & -\,\sigma_{-+}\\[2ex] - \,\sigma_{+-} & \sigma_{-+}\end{array}\right] \ \left[\begin{array}{c} N_+ \\[2ex] N_-\end{array}\right] \, ,
\end{eqnarray}
{\it i.e.}\ 
\begin{eqnarray}
\label{eq:N+N-No_Loss_System3}
\frac{d\,N_+}{d\,\tau} & = & - \,n\,\nu \ \left(\,\sigma_{+-}\,N_+ \ - \ \sigma_{-+}\,N_-\,\right)  \, ,\nonumber \\[2ex]
\frac{d\,N_-}{d\,\tau} & = & - \,n\,\nu \ \left(\,-\,\sigma_{+-}\,N_+ \ + \ \sigma_{-+}\,N_-\,\right) \, ,
\end{eqnarray}
and adding these gives
\begin{eqnarray*}
\frac{d\,N}{d\,\tau} & = & - \,n\,\nu \ \left(\,\sigma_{+-}\,N_+ \ - \ \sigma_{-+}\,N_- \ - \ \sigma_{+-}\,N_+ \ + \ \sigma_{-+}\,N_-\,\right) \ \,=\, \ 0 \, .
\end{eqnarray*}
So we have $d\,/\,d\,\tau \ \left[\,N_+(\tau) \ + \ N_-(\tau)\,\right] \,=\, 0$ which implies $N_+(\tau) \ + \ N_-(\tau) \,=\, \mbox{constant} \,=\, N_0$\,.  Thus there will be no loss of particles, as expected.  Subtracting eqs.~(\ref{eq:N+N-No_Loss_System3}) from each other gives
\begin{eqnarray*}
\frac{d\,J}{d\,\tau} & = & - \,2\,n\,\nu \ \left(\,\sigma_{+-}\,N_+ \ - \ \sigma_{-+}\,N_-\,\right).
\end{eqnarray*}
which leads to a non-zero $J(\tau)$ ({\it i.e.}\ non-zero polarization) provided that $\sigma_{+-} \ \neq \ \sigma_{-+}$\,.

This point is highlighted in the Th.~Walcher {\it et al.} paper \cite{Walcher:2007sj}, where for a co-moving positron beam the maximum antiproton scattering angle is less than the ring acceptance angle, thus no particles are scattered out of the beam.  In fact eqs.~(\ref{eq:N+N-No_Loss_System3}) above are identical to eqs.~(45 and 46) of ref.~\cite{Walcher:2007sj}, upon which Th.~Walcher {\it et al.} base their dynamics.

The system without scattering out of the beam described in eq.~(\ref{eq:N+N-No_Loss_System}) is very similar to the system which describes the Sokolov-Ternov effect of radiative polarization \cite{Krisch:1986nt,Sokolov:1963,Lee:1997,Leader:2005}.  The systems describing these two physical processes should be similar as they are both governed solely by spin-flip transitions.  In spin filtering the spin-flip transitions are induced by scattering off the polarized internal target while in the Sokolov-Ternov effect the spin-flip transitions are induced by spontaneous synchrotron radiation emission of photons while the charged particles of the beam are being bent in the magnetic field of the ring.  In fact these systems are identical except for the interpretations of the matrix entries, and that because there is no target in the Sokolov-Ternov effect, the system of equations describing it does not depend on a target areal density $n$.  While electrons and positrons circulating in a storage ring acquire a useful polarization in a relatively short time due to the Sokolov-Ternov effect \cite{Lee:1997,Leader:2005}, an unrealistically large time would be required in the case of protons or antiprotons at currently achievable energies, due to their high mass \cite{Krisch:1986nt}.  While during spin filtering in a storage ring polarization buildup due to the Sokolov-Ternov effect does happen, for low energy antiprotons the rate is many orders of magnitude lower than the rate of polarization buildup due to spin filtering \cite{Krisch:1986nt}.  Thus the Sokolov-Ternov effect may be neglected in the treatment of antiproton polarization buildup by spin filtering, at the low energies of interest.
\\
\\
{\bf 3)} {\it When there is no scattering out of the beam, {\it i.e.}\ $\sigma^\mathrm{\,out}_+ \, = \, \sigma^\mathrm{\,out}_- \, = \,0$, the condition
\begin{equation}
\label{eq:No_Loss_Conservation_of_N}
\frac{d\,N_+}{d\,\tau} \ = \ - \, \frac{d\,N_-}{d\,\tau} \, ,
\end{equation}
must be satisfied.}
\\
We have shown that when there is no scattering out of the beam the polarization evolution equations reduce to eqs.~(\ref{eq:N+N-No_Loss_System3}) which can immediately be seen to satisfy eq.~(\ref{eq:No_Loss_Conservation_of_N}).

This point is highlighted in the Th.~Walcher {\it et al.} paper \cite{Walcher:2007sj}, where for a co-moving polarized positron beam there is no scattering out of the beam.
\\
\\
{\bf 4)} {\it When there is no spin-flip the change in one spin state should not depend on the number of particles in the other spin state, {\it i.e.}\ the equations should decouple.}
\\
This can be shown as follows: When there is no spin-flip $\sigma_{+-}\, = \, \sigma_{-+}\, = \,0$ and thus the polarization evolution equations presented in eqs.~(\ref{eq:N+N-Polarization_evolution_Equations}) reduce to
\begin{eqnarray}
\label{eq:No_Spin_Flip_System}
\frac{d\,N_+}{d\,\tau} & = & -\,n\,\nu\,\sigma^\mathrm{\,out}_+ \,N_+  \, ,\nonumber \\[2ex]
 \frac{d\,N_-}{d\,\tau} & = & -\,n\,\nu\,\sigma^\mathrm{\,out}_- \,N_- \, ,
\end{eqnarray}
which is an uncoupled system of equations as required.

This point is highlighted in W.~W.~MacKay and C.~Montag's paper \cite{MacKay:2006}.  One sees that if in addition $\sigma^\mathrm{\,out}_+ \,=\,\sigma^\mathrm{\,out}_-$ in eqs.~(\ref{eq:No_Spin_Flip_System}) then no polarization buildup occurs.  It is claimed by the Budker-J\"ulich groups that the spin-flip transition rates are negligible for antiprotons scattering off polarized electrons in a hydrogen target \cite{Milstein:2005bx,Nikolaev:2006gw}.  The maximum scattering angle of antiprotons scattering off atomic electrons is $0.54 \ \mbox{mrad}$ \cite{Milstein:2005bx}, below the acceptance angle of a typical storage ring, thus there is no scattering out of the beam.  Since there is no scattering out of the beam, and spin-flip transitions are negligible the Budker-J\"ulich groups conclude that polarized electrons in an atomic target are not effective in transferring polarization to an antiproton beam by spin filtering \cite{Milstein:2005bx,Nikolaev:2006gw}.  To force some antiprotons to be scattered out of the beam, and to avoid the problem of loss of beam intensity due to antiprotons annihilating with the protons in an atomic target, it has been suggested to use an opposing polarized electron beam of sufficient energy to scatter some antiprotons beyond the ring acceptance angle \cite{O'Brien:2007hu}.  The rate of polarization buildup using this method is slow due to the low densities of polarized electron beams currently available \cite{O'Brien:2007hu}, but the enhanced cross-sections at low energies suggested in refs.~\cite{Walcher:2007sj,Arenhoevel:2007} may compensate this difficulty.

\section{$\sigma^\mathrm{\,out}_+$\,,\, $\sigma^\mathrm{\,out}_-$\,,\, $\sigma_{+-}$\,,\,$\sigma_{-+}$ and the spin observables}
\label{sec:Relations_between_cross-sections_and_spin_observables}

The spin observables of a spin 1/2 - spin 1/2 scattering process are defined in refs.~\cite{Bystricky:1976jr,LaFrance:1980}.  In spin filtering where the polarization of the recoiled target particle is not important one is interested in the polarization transfer, depolarization and double spin asymmetry spin observables.  These have been calculated for electromagnetic antiproton-proton and antiproton-electron elastic scattering in ref.~\cite{O'Brien:2006zt}.  The spin transfer observable has been calculated for low energy antiproton-positron scattering in ref.~\cite{Arenhoevel:2007}.  A large increase of this spin transfer cross-section at very low energies is the basis for the proposal to polarize antiprotons by interaction with a co-moving polarized positron beam presented in ref.~\cite{Walcher:2007sj}.

The cross-sections $\sigma^\mathrm{\,out}_+$, $\sigma^\mathrm{\,out}_-$, $\sigma_{+-}$ and $\sigma_{-+}$ can be related to the spin observables by the following relations \cite{Nikolaev:2006gw}:
\begin{eqnarray}
\label{eq:N+N-_to_N_J_Relation1}
\sigma^\mathrm{\,out}_+ & \equiv &  I_\mathrm{\,out} \ +\  P_T \ A_\mathrm{\,out}  \, ,\\[2ex]
\label{eq:N+N-_to_N_J_Relation2}
\sigma^\mathrm{\,out}_- & \equiv & I_\mathrm{\,out} \ - \ P_T \ A_\mathrm{\,out}  \, ,\\[2ex]
\label{eq:N+N-_to_N_J_Relation3}
\sigma_{+-}    & \equiv & L_\mathrm{\,in} \ + \  \frac{P_T}{2} \ \left(\,A_\mathrm{\,in} \ - \ K_\mathrm{\,in}\,\right)  \, ,\\[2ex]
\label{eq:N+N-_to_N_J_Relation4}
\sigma_{-+}    & \equiv & L_\mathrm{\,in} \ - \  \frac{P_T}{2} \ \left(\,A_\mathrm{\,in} \ -\  K_\mathrm{\,in}\,\right)  \, ,
\end{eqnarray}
where $P_T$ is the polarization of the target, and $L_\mathrm{\,in} \, = \, \left(\,I_\mathrm{\,in} - D_\mathrm{\,in}\,\right)\,/\,2$ is a loss of polarization quantity.  $I = {\mathrm{d}}\sigma\,/\,{\mathrm{d}}\Omega$ is the spin averaged differential cross-section and $A$, $K$ and $D$ are the double spin asymmetry, polarization transfer and depolarization spin observables respectively as calculated in ref.~\cite{O'Brien:2006zt}.  As we show later, the relations above are crucial for comparing the different theoretical descriptions of spin filtering which appear in the literature.  The subscripts {\bf ``in''}, {\bf ``out''} and {\bf ``all''} refer to angular integration of the spin observables over the following ranges:  The {\bf``in''} subscript refers to particles that are scattered at small angles $\leq \theta_\mathrm{\,acc}$ remaining in the beam, and the {\bf``out''} subscript refers to particles that are scattered out of the beam.  Thus the integrals over scattering angle $\theta$ are labeled {\bf``in''} where the range of integration is $\theta_\mathrm{\,min} \leq \theta \leq \theta_\mathrm{\,acc}$, {\bf``out''} where the range of integration is $\theta_\mathrm{\,acc} < \theta \leq \pi$ and {\bf``all''} $=$ {\bf``in''} $+$ {\bf``out''} where the range of integration is $\theta_\mathrm{\,min} \leq \theta \leq \pi$ as seen in table~1 of ref.~\cite{Buttimore:2007cj}. 

Note the following linear combinations of the cross-sections
\begin{eqnarray}
I_\mathrm{\,out} & = &  \frac{\sigma^\mathrm{\,out}_+ \ + \  \sigma^\mathrm{\,out}_-}{2}  \, ,\\[2ex]
\label{eq:Loss_cross_section_difference}
P_T\,A_\mathrm{\,out} & = & \frac{\sigma^\mathrm{\,out}_+ \ - \ \sigma^\mathrm{\,out}_-}{2}  \, ,\\[2ex]
L_\mathrm{\,in} & = &  \frac{\sigma_{+-} \ + \  \sigma_{-+}}{2}  \, ,\\[2ex]
\label{eq:Spin_flip_cross_section_difference}
P_T \ \left(\,A_\mathrm{\,in} \ - \  K_\mathrm{\,in}\,\right) & = & \sigma_{+-} \ - \ \sigma_{-+} \, .
\end{eqnarray}
Again to ensure consistency and to highlight the physical properties we are trying to describe mathematically we prove three short results on the above relations between the cross-sections and the spin observables.
\\
\\
{\bf 1)} {\it If the target is unpolarized ($P_T \,=\, 0$) then one has that $\sigma^\mathrm{\,out}_+ \,=\, \sigma^\mathrm{\,out}_-$ and $\sigma_{+-} \,=\, \sigma_{-+}$ so no polarization buildup will occur.}
\\
This can be shown as follows: Setting $P_T \,=\, 0$ into the eqs.~(\ref{eq:N+N-_to_N_J_Relation1}, \ref{eq:N+N-_to_N_J_Relation2}, \ref{eq:N+N-_to_N_J_Relation3} and \ref{eq:N+N-_to_N_J_Relation4}) one immediately obtains $\sigma^\mathrm{\,out}_+ \, = \,  I_\mathrm{\,out}  \, = \, \sigma^\mathrm{\,out}_-$ and $\sigma_{+-} \, = \, L_\mathrm{\,in}  \, = \,\sigma_{-+}$.  Once this is satisfied it has been proved earlier, using eq.~(\ref{eq:System_with_Unpolarized_target2}), that no polarization buildup will occur in this case.
\\
\\
{\bf 2)} {\it The spin-flip cross-sections should depend only on spin observables relating to particles scattering within the ring, {\it i.e.}\ only to {\bf ``in''} subscripted spin observables which are integrated from $\theta_\mathrm{\,min}$ to $\theta_\mathrm{\,acc}$\,.}
\\
This is immediately satisfied by the relations in eqs.~(\ref{eq:N+N-_to_N_J_Relation3} and \ref{eq:N+N-_to_N_J_Relation4}).
\\
\\
{\bf 3)} {\it The cross-section differences $\sigma^\mathrm{\,out}_+ \, - \, \sigma^\mathrm{\,out}_-$ and $\sigma_{+-} \, - \,\sigma_{-+}$ should both be proportional to the target polarization $P_T$.}
\\
This is immediately satisfied by the relations in eqs.~(\ref{eq:Loss_cross_section_difference} and \ref{eq:Spin_flip_cross_section_difference}).
\\
\\
While the system of polarization evolution equations involving the variables $N_+(\tau)$ and $N_-(\tau)$ presented in eq.~(\ref{eq:N+N-Polarization_evolution_Equations}) is very transparent, one is more interested in the variables $N(\tau)$ and $J(\tau)$ which immediately lead to $P(\tau) = J(\tau)\,/\,N(\tau)$.  We can transform the system of two first order coupled ODE in variables $N_+(\tau)$ and $N_-(\tau)$ presented in eq.~(\ref{eq:N+N-Polarization_evolution_Equations}) to the following system of two first order ODE in variables $N(\tau)$ and $J(\tau)$ \cite{Nikolaev:2006gw,O'Brien:2007hu,Buttimore:2007cj}\,:
\begin{eqnarray}
\label{eq:HomogeneousSystem}
  \frac{\mathrm{d}}{\mathrm{d}\tau}
\left[
        \begin{array}{c} N \\[2ex] J \end {array}
\right]
\,  = \, - \, n \, \nu
\left[
\begin{array}{ccc}
         I_\mathrm{\, out} && P_T \, A_\mathrm{\, out}
\\[2ex]
    P_T \left(\,A_\mathrm{\, all} \, - \, K_\mathrm{\,in}\,\right)
&&
         I_\mathrm{\, all} \, - \,   D_\mathrm{\, in}
\end {array}
\right]
\,
\left[
        \begin{array}{c} N \\[2ex] J \end {array}
\right] \ ,
\end{eqnarray}
where we have also transformed from the cross-sections to the spin observables which have already been calculated \cite{O'Brien:2006zt}.  The above system of polarization evolution equations has been presented by N.~N.~Nikolaev and F.~F.~Pavlov \cite{Nikolaev:2006gw} and is further discussed in refs.~\cite{Buttimore:2007cj,O'Brien:2007sw}.  The systems presented in eqs.~(\ref{eq:N+N-Polarization_evolution_Equations}) and eqs.~(\ref{eq:HomogeneousSystem}) are identical provided eqs.~(12--15) hold.  From now on we concentrate on the latter as its solution is more illustrative of the underlying physical phenomena, and the dependence on the target polarization is explicit.  In particular one immediately sees that when the target is unpolarized no beam polarization buildup occurs, as when $P_T \,=\, 0$ the system reduces to two uncoupled separable first order ODE as in eq.~(\ref{eq:System_with_Unpolarized_target1}) with solutions as presented in eq.~(\ref{eq:System_with_Unpolarized_target2}) showing $P(\tau) \,=\, 0 \ \mbox{for all} \ \tau \ \mbox{if} \ P_T \,=\, 0$.

\section{Solution of the polarization evolution equations}
\label{sec:Solution_of_the_polarization_evolution_equations}

In this section we solve the polarization evolution equations presented in eq.~(\ref{eq:HomogeneousSystem}).  The eigenvalues of the matrix of coefficients are found to be 
\begin{equation}
\lambda_1 \ = \ - \,n\,\nu\, \left(\,I_\mathrm{\,out} \,+\, L_\mathrm{\,in} \,+\, L_\mathrm{\,d}\,\right) 
\hspace*{1em} \mbox{and} \hspace*{1em}
\lambda_2 \ = \ - \,n\,\nu\, \left(\,I_\mathrm{\,out} \,+\, L_\mathrm{\,in} \,-\, L_\mathrm{\,d}\,\right) \, ,
\end{equation}
where the discriminant $L_\mathrm{\,d}$ of the quadratic equation
 for the eigenvalues is
\begin{equation}
   L_\mathrm{\,d}
\ =
\ \sqrt{\, P_T^{\,2} \, A_\mathrm{\,out} \left( A_\mathrm{\,all}
\, -
\, K_\mathrm{\,in} \right) \, + \, L_\mathrm{\,in}^{\,2} }  \ \, .
\end{equation}
Note that $I_\mathrm{\,out}$, $L_\mathrm{\,in}$ and $L_\mathrm{\,d}$ are all positive.  As a consequence the eigenvalues are negative and $\lambda_1 < \lambda_2 < 0$.

Now enforcing the initial conditions $N(0) \,=\, N_0$ the total number of particles in the beam initially, and $J(0) \,=\, 0 \Rightarrow N_+(0) \,=\, N_-(0) \,=\, N_0\,/\,2$ {\it i.e.}\ initially the beam is unpolarized, one obtains the solutions:
\begin{eqnarray}
\label{eq:Homogeneous_N}
N(\tau) & = & \frac{\left[\,e^{\,\lambda_1\,\tau}\, \left(\,L_\mathrm{\,d} - L_\mathrm{\,in}\,\right) \, + \, e^{\,\lambda_2\,\tau}\, \left(L_\mathrm{\,d} + L_\mathrm{\,in}\right)\,\right]\,N_0}{2\, L_\mathrm{\,d}} \, ,\\[2ex]
\label{eq:Homogeneous_J}
J(\tau) & = & \displaystyle{ \frac{\left(\,e^{\,\lambda_1\,\tau}-e^{\,\lambda_2\,\tau}\,\right)\,\left(A_\mathrm{\,all} - K_\mathrm{\,in}\right)\,N_0\,P_T}{2\, L_\mathrm{\,d}}}\,. 
\end{eqnarray}
The time ($\tau$) dependence of the polarization of the beam is given by
\begin{eqnarray}
\label{eq:HomogeneousPolarizationBuildup}
P(\tau) \ \,=\, \ \frac{J(\tau)}{N(\tau)} \ \,=\, \ \frac{ -\,P_T\,\left(\,A_\mathrm{\,all} \, - \, K_\mathrm{\,in}\,\right)}{L_\mathrm{\,in} \, + \, L_\mathrm{\,d} \, \coth\left(L_\mathrm{\,d}\, n\,\nu\,\tau\right)} \, .
\end{eqnarray}
The expression for $P(\tau)$ is proportional to $P_T$ which confirms that if the target polarization is zero there will be no polarization buildup in the beam.
The approximate rate of change of
 polarization for sufficiently short times, and the limit of the polarization for large times are respectively:
\begin{equation}
\label{eq:Homogeneous_initial_and_max_polarization}
   \frac{\mathrm{d}\,P}{\mathrm{d}\tau} \ \approx \ -\, n \, \nu \, P_T
\,
\left( A_\mathrm{\,all} \, - \, K_\mathrm{\,in} \right)
\hspace*{1em} \mbox{and} \hspace*{1em}
\displaystyle{\lim_{\tau \to \,\infty} P(\tau) \ = \ -\, P_T
\,
\frac{ A_\mathrm{\,all} \, - \, K_\mathrm{\,in}
}
{   L_\mathrm{\,in} \, + \, L_\mathrm{\,d}
}}
\,.
\end{equation}
For pure electromagnetic scattering the double spin asymmetries equal the polarization transfer spin observables \cite{O'Brien:2006zt}, thus one can simplify the above equations using $A_\mathrm{\,in} \,=\, K_\mathrm{\,in}$, $A_\mathrm{\,out} \,=\, K_\mathrm{\,out}$ and $A_\mathrm{\,all} \,=\, K_\mathrm{\,all}$\,; hence $A_\mathrm{\,all} \,-\, K_\mathrm{\,in} \,=\, K_\mathrm{\,out}$\,.  

The solution to the above system and to many alterations to this system accounting for spin filtering in various scenarios have been presented recently \cite{Buttimore:2007cj,O'Brien:2007sw}.  These scenarios are: 1) spin filtering of a fully stored beam, 2) spin filtering while the beam is being accumulated, {\it i.e.}\ unpolarized particles are continuously being fed into the beam at a constant rate, 3)
unpolarized particles are continuously being fed into the beam at a linearly increasing rate, {\it i.e.}\ the particle input rate is ramped up, 4) the particle input rate is equal to the rate at which particles are being lost due to scattering beyond ring acceptance angle, the beam intensity remaining constant, 5) increasing the initial polarization of a stored beam by spin filtering, and 6) the input of particles into the beam is stopped after a certain amount of time, but spin filtering continues.


\begin{theacknowledgments}
This work was kindly funded by the Irish Research Council for Science Engineering and Technology (IRCSET).  I would also like to thank M.~Anselmino, N.~H.~Buttimore, E.~Leader, H.~O.~Meyer, N.~N.~Nikolaev, F.~Rathmann, G.~Stancari and Th.~Walcher for helpful comments.
\end{theacknowledgments}


\end{document}